\documentclass{jfm}
\usepackage[utf8]{inputenc}

\usepackage{amsmath,amssymb,mathtools}
\usepackage{natbib}
\usepackage{overpic}
\usepackage{xcolor}
\usepackage{hyperref}
\usepackage{cleveref}
\allowdisplaybreaks

\title{Emergent rheotaxis of shape-changing swimmers in Poiseuille flow}
\date{\today}

\shorttitle{Emergent rheotaxis of shape-changing swimmers in Poiseuille flow}
\shortauthor{Walker, Ishimoto, Moreau, Gaffney, and Dalwadi}
\author{B. J. Walker\aff{1\corresp{\email{benjamin.walker@ucl.ac.uk}}}, K. Ishimoto\aff{2}, C. Moreau\aff{2}, E. A. Gaffney\aff{3}, M. P. Dalwadi\aff{1}}

\affiliation{\aff{1}Department of Mathematics, University College London, London, WC1H 0AY, UK
\aff{2}Research Institute for Mathematical Sciences, Kyoto University, Kyoto, 606-8502, Japan
\aff{3}Wolfson Centre for Mathematical Biology, Mathematical Institute, University of Oxford, Oxford, OX2 6GG, UK}

\newcommand{\bigO}[1]{O\left(#1\right)}
\newcommand{\diff}[2]{\frac{\mathrm{d}#1}{\mathrm{d}#2}}
\newcommand{\pdiff}[2]{\frac{\partial #1}{\partial #2}}
\newcommand{\avg}[1]{\left\langle #1 \right\rangle}
\newcommand{\intd}{\mathop{}\!\mathrm{d}}
\newcommand{\dummy}[1]{\tilde{#1}}
\newcommand{\Iu}{I_u}
\newcommand{\IB}{I_B}
\DeclareMathOperator\arctanh{arctanh}
\newcommand{\tauperiod}{P_{\tau}}
\newcommand{\abs}[1]{\left\vert #1 \right\vert}

\begin{document}
    
\maketitle

\begin{abstract}
A simple model for the motion of shape-changing swimmers in Poiseuille flow was recently proposed and numerically explored by \citet{Omori2022}. These explorations hinted that a small number of interacting mechanics can drive long-time behaviours in this model, cast in the context of the well-studied alga \textit{Chlamydomonas} and its rheotactic behaviours in such flows. Here, we explore this model analytically via a multiple-scale asymptotic analysis, seeking to formally identify the causal factors that shape the behaviour of these swimmers in Poiseuille flow. By capturing the evolution of a Hamiltonian-like quantity, we reveal the origins of the long-term drift in a single swimmer-dependent constant, whose sign determines the eventual behaviour of the swimmer. This constant captures the nonlinear interaction between the oscillatory speed and effective hydrodynamic shape of deforming swimmers, driving drift either towards or away from rheotaxis.
\end{abstract}

\section{Introduction}
\begin{figure}
    \centering
    \includegraphics[width=0.5\textwidth]{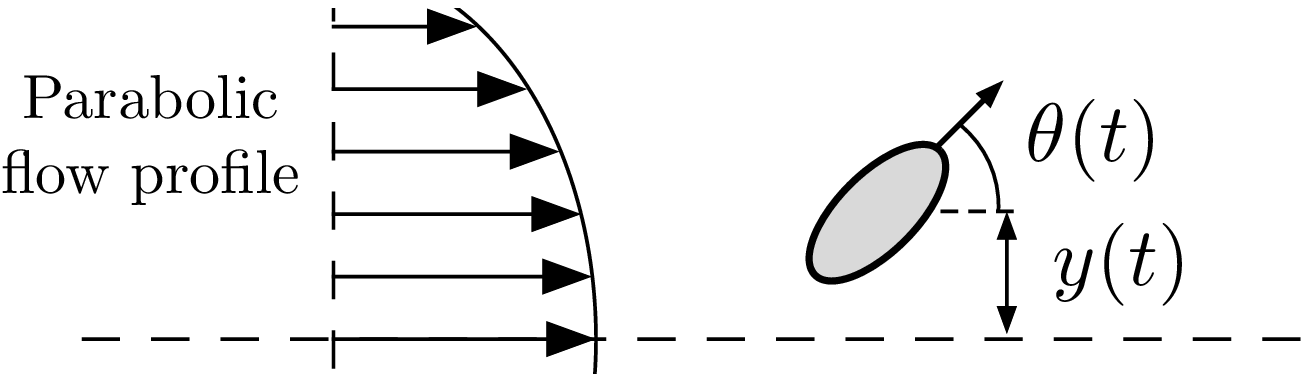}
    \caption{Notation and set-up. We illustrate a model swimmer in Poiseuille flow, located at a transverse displacement $y$ from the midline of the parabolic flow profile. The swimming direction $\theta$ is measured from the midline, with $\theta=0$ corresponding to downstream swimming.}
    \label{fig:setup}
\end{figure}

The behaviours of microswimmers in flows have long been a topic of broad theoretical and experimental study. Recently, \citet{Omori2022} numerically explored a model of a shape-changing swimmer in Poiseuille flow, posed in the context of the alga \textit{Chlamydomonas} for comparison  with their experimental findings. Their investigations suggested an interesting and subtle connection between the long-time behaviours of the microswimmer and the details of its changing speed and shape, with certain conditions apparently necessary for long-time upstream-facing swimming in the flow, referred to as \emph{rheotaxis} by \citeauthor{Omori2022}. Their ordinary differential equation (ODE) model may be simply stated in terms of a transverse coordinate $y$ and the swimmer orientation $\theta$ as\begin{subequations}\label{eq: original system}
\begin{align}
    \diff{y}{t} &= \omega u(\omega t)\sin{\theta}\,,\label{eq: original system: y}\\
    \diff{\theta}{t} &= \gamma y (1 - B(\omega t)\cos{2\theta})\,,\label{eq: original system: theta}
\end{align}
\end{subequations}
with given initial conditions and with reference to the setup of \cref{fig:setup}, where all quantities are considered dimensionless. We refer the interested reader to the work of \citet{Omori2022} for a full derivation of the model. The functions $u$ and $B$ capture the time-dependent active swimming speed and shape-capturing Bretherton constant \citep{Bretherton1962}, respectively. These prescribed functions are assumed to be oscillatory with a shared period of unity in $\omega t$, where the parameter $\omega\gg1$ encodes the high frequency of these oscillations. The swimming speed naturally scales with $\omega$ in \cref{eq: original system: y}, though we note that such a scaling is absent from the explicitly stated equations of \citet{Omori2022} but is present in their explored parameter regimes. Here, $\gamma$ is a fixed characteristic shear rate of the flow, non-negative without loss of generality. This model neglects any interactions of the swimmer with solid boundaries typically associated with Poiseuille flow, and we will proceed without additional consideration of boundary effects.
 
Via the numerical explorations of \citet{Omori2022}, this model is noted to give rise to a range of complex, long-time behaviours, perhaps most remarkable of which is conditional convergence towards a central upstream-facing configuration. In this study, we will aim to analytically uncover the driving factors behind these long-time dynamics. Via a multiple-scale asymptotic analysis \citep{Bender1999}, as recently applied to similar models of swimming \citep{Walker2022,Gaffney2022,Ma2022}, we will show how the effective swimmer behaviour can be captured by a Hamiltonian-like quantity, whose slow evolution accurately encodes the long-time trends of behaviour noted by \citet{Omori2022}. Further, we will identify a markedly simple relation between the eventual behaviour of the swimmer and its oscillating speed and shape, enabling the deduction of long-time dynamics through the calculation of a single swimmer-dependent constant.
 

\section{Direct asymptotic analysis}
The timescales present in the model of \citet{Omori2022} are best identified through a change of variable. Defining $z(t)\coloneqq y(t)/\omega^{1/2}$, the system reads\begin{subequations}\label{eq: z-theta system}
\begin{align}
    \diff{z}{t} &=  \omega^{1/2}u(\omega t)\sin{\theta}\,,\label{eq: z-theta system: z}\\
    \diff{\theta}{t} &= \gamma \omega^{1/2}z (1 - B(\omega t)\cos{2\theta})\,.\label{eq: z-theta system: theta}
\end{align}
\end{subequations}
This suggests a natural fast timescale $T\coloneqq \omega t$, that of the oscillating swimmer speed and shape. Additionally, $\bigO{1}$ oscillations of $u$ and $B$ in \cref{eq: z-theta system} drive $\bigO{1}$ changes in $z$ and $\theta$ over an intermediate timescale of $t=\bigO{\omega^{-1/2}}$. We will later see that these changes correspond to quasiperiodic orbits of the swimmer in the flow, quantifying motion on this timescale via $\tau\coloneqq \omega^{1/2}t$. In addition to the two timescales $T$ and $\tau$ evident from this system of equations, \citet{Omori2022} observed behavioural changes over longer timescales, with $t=\bigO{1}$. Hence, we expect the system to evolve on three separated timescales, corresponding to $T$, $\tau$, and $t$ each being $\bigO{1}$. Our overarching aim is to characterise and understand the behaviours of the system over the timescale $t = \bigO{1}$.

To this end, we implement a multiple-scale analysis and formally write $z(t) = z(T,\tau,t)$ and $\theta(t) = \theta(T,\tau,t)$, treating each time variable as independent \citep{Bender1999}. This transforms the proper time derivative via 
\begin{equation}\label{eq: transform derivative}
    \diff{}{t} \mapsto \omega\pdiff{}{T} + \omega^{1/2}\pdiff{}{\tau} + \pdiff{}{t}\,,
\end{equation}
which transforms \cref{eq: z-theta system} into the system of partial differential equations (PDEs)\begin{subequations}\label{eq: z-theta MS system}
\begin{align}
    \omega z_T + \omega^{1/2}z_{\tau} + z_t &= \omega^{1/2} u(T)\sin{\theta}\,,\label{eq: z-theta MS system: z}\\
    \omega\theta_{T} + \omega^{1/2}\theta_{\tau} + \theta_{t} &= \omega^{1/2}\gamma z (1 - B(T)\cos{2\theta})\,.\label{eq: z-theta MS system: theta}
\end{align}
\end{subequations}
Here and hereafter, subscripts of $t$, $\tau$, and $T$ denote partial derivatives. We will later remove the extra degrees of freedom that we have introduced by imposing periodicity of the dynamics in the intermediate and fast variables $\tau$ and $T$. Expanding $z$ and $\theta$ in powers of $\omega^{-1/2}$ as $z \sim z_0 + \omega^{-1/2}z_1 + \omega^{-1}z_2 + \cdots$ and $\theta \sim \theta_0 + \omega^{-1/2}\theta_1 + \omega^{-1}\theta_2 + \cdots$, we obtain the $\bigO{\omega}$ balance
\begin{equation}\label{eq: zeroth order balance}
    z_{0T} = 0\,, \quad \theta_{0T} = 0\,,
\end{equation}
so that $z_0=z_0(\tau,t)$ and $\theta_0=\theta_0(\tau,t)$ are independent of $T$. To determine how $z_0$ and $\theta_0$ depend on $\tau$ and $t$, we must proceed to higher asymptotic orders. 

We next consider the balance of $\bigO{\omega^{1/2}}$ terms in \cref{eq: z-theta MS system}, which reads\begin{subequations}\label{eq: first order balance}
\begin{align}
    z_{1T} + z_{0\tau} & = u(T)\sin{\theta_0}\,,\\
    \theta_{1T} + \theta_{0\tau} & = \gamma z_0(1 - B(T)\cos{2\theta_0})\,.
\end{align}
\end{subequations}
The Fredholm solvability condition for \cref{eq: first order balance} is equivalent to averaging over a period in $T$ and enforcing $T$-periodicity of $z_1$ and $\theta_1$. Introducing the averaging operator $\avg{\cdot}$, defined via its action on functions $v(T,\tau,t)$ via
\begin{equation}
    \avg{v}(\tau,t) \coloneqq \int_0^1 v(T,\tau,t)\intd{T}\,,
\end{equation}
we obtain the averaged equations\begin{subequations}\label{eq: first order solvability}
\begin{align}
    z_{0\tau} & = \avg{u}\sin{\theta_0}\,,\label{eq: first order solvability: z0}\\
    \theta_{0\tau} & = \gamma z_0(1 - \avg{B}\cos{2\theta_0})\,,
\end{align}
\end{subequations}
where $\avg{u}$ and $\avg{B}$ are the averages of $u(T)$ and $B(T)$, respectively, representing the average speed and shape of the model swimmer. In particular, $\avg{u}$ and $\avg{B}$ are constant, with the dynamics being rendered trivial if $\avg{u}=0$; we exclude this case from our analysis and henceforth take $\avg{u}>0$ without further loss of generality\footnote{The mapping $\theta\mapsto\theta+\pi$ transforms $\avg{u}<0$ into the positive case.}. We will also assume that $\abs{\avg{B}}<1$, which imposes only a minimal restriction on the admissible swimmer shapes, since $\abs{B}\geq1$ is  typically associated with objects of exceedingly large aspect ratio \citep{Bretherton1962}.

Of particular note, if viewed as a system of ODEs in $\tau$, the system of \cref{eq: first order solvability} corresponds precisely to the original dynamical system of \cref{eq: z-theta system}, suitably scaled, but with the time-varying speed and shape parameters replaced by their averages. We will shortly return to these equations and explore the ramifications of this observation in detail, in particular noting the existence of a Hamiltonian-like quantity, but first complete our analysis of the $\bigO{\omega^{1/2}}$ problem to determine the form of $z_1$ and $\theta_1$ for later convenience.

Without solving \cref{eq: first order solvability}, we can deduce the form of $z_1$ and $\theta_1$ by substituting \cref{eq: first order solvability} into \cref{eq: first order balance}, yielding the simplified system\begin{subequations}\label{eq: solvable z1-theta1 system}
\begin{align}
    z_{1T} & = [u(T) - \avg{u}]\sin{\theta_0}\,,\\
    \theta_{1T} & = -\gamma z_0[B(T) - \avg{B}]\cos{2\theta_0}\,.
\end{align}
\end{subequations}
Integrating \cref{eq: solvable z1-theta1 system} in $T$, recalling that $z_0$ and $\theta_0$ are independent of $T$, yields the solution\begin{subequations}\label{eq: z1-theta1 solution}
\begin{align}
    z_1 &= \Iu(T)\sin{\theta_0} + \bar{z}_1(\tau,t)\,,\\
    \theta_1 & = -\gamma z_0\IB(T)\cos{2\theta_0} + \bar{\theta}_1(\tau,t)\,,
\end{align}
\end{subequations}
where $\bar{z}_1$ and $\bar{\theta}_1$ are functions of $\tau$ and $t$, undetermined at this order, and we define
\begin{equation}
    \Iu(T) \coloneqq \int_0^T [u(\dummy{T}) - \avg{u}] \intd{\dummy{T}}\,, \quad \IB(T) \coloneqq \int_0^T [B(\dummy{T}) - \avg{B}] \intd{\dummy{T}}\,.
\end{equation}
Noting that $\Iu(T)$ and $\IB(T)$ are $T$-periodic with period one, it follows that $z_1$ and $\theta_1$ are $T$-periodic with the same period.

In principle, one could proceed to the next asymptotic order to determine how $z_0$ and $\theta_0$ evolve in $t$ through the derivation of an additional solvability condition. However, here, this procedure would be complicated by the absence of an explicit solution to the nonlinear system of \cref{eq: first order solvability}, compounded by the potentially $t$-dependent period of the solution, which would require using the generalised method of \citet{Kuzmak1959}. To circumvent this difficulty, we instead turn our attention back to \cref{eq: first order solvability}, seeking further understanding of the leading-order dynamics over the intermediate timescale $\tau$.

\section{A Hamiltonian-like quantity}
\begin{figure}
    \centering
    \includegraphics[width=0.45\textwidth]{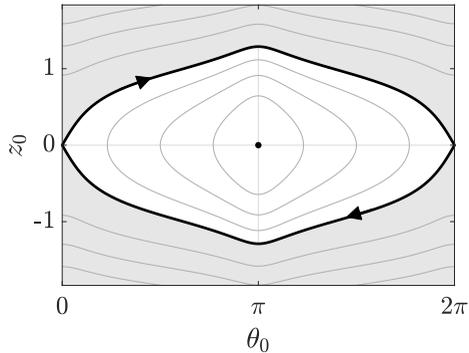}
    \caption{Phase portrait of motion on the intermediate timescale $\tau$. Solutions of \cref{eq: H_0(t)} are closed orbits in the $z_0$-$\theta_0$ plane for constant $H_0$, symmetric in both $z_0=0$ and $\theta_0=\pi$. Solutions in the shaded region, where $H_0 > g(0)$, do not cross $z_0=0$, corresponding to tumbling motion and monotonic evolution of $\theta_0$. Trajectories with $H_0 < g(0)$ instead exhibit swinging motion, with $\theta_0$ oscillating between two values. The black contour $H_0 = g(0)$ separates these regimes, with the direction of motion in the phase plane indicated by arrowheads, recalling that $\gamma\geq0$. The point $(z_0,\theta_0)=(0,\pi)$ corresponds to rheotaxis, with $H_0=g(\pi)$.}
    \label{fig: phase space}
\end{figure}

If treated as a system of ODEs, we noted that \cref{eq: first order solvability} closely resembles the original swimming problem, with averages taking the place of oscillatory swimming speeds and swimmer shapes. In fact, the equivalent ODE problem has been extensively studied, with \citet{Zottl2013a} thoroughly exploring this dynamical system and identifying a Hamiltonian-like constant of motion. Motivated by their study, we identify an analogous first integral of \cref{eq: first order solvability}:
\begin{equation}\label{eq: H_0(t)}
   H_0(t) \coloneqq  \frac{\gamma}{2\avg{u}}z_0^2 + g(\theta_0)
\end{equation}
for $H_0(t)\in[g(\pi),\infty)$, where, for $\avg{B}\in(-1,1)$,
\begin{equation}
    g(\theta_0) \coloneqq \frac{\arctanh{\left(\sqrt{\frac{2\avg{B}}{1+\avg{B}}}\cos{\theta_0}\right)}}{\sqrt{2\avg{B}(1+\avg{B})}}\,, \quad g'(\theta_0) = -\frac{\sin{\theta_0}}{1 - \avg{B}\cos{2\theta_0}}\,,
\end{equation}
taking the appropriate limits and branches where required. As $H_0(t)$ is effectively a constant of motion over the intermediate timescale $\tau$, \cref{eq: H_0(t)} demonstrates that solutions to \cref{eq: first order solvability} are closed orbits in $z_0$-$\theta_0$ phase space over this timescale, with $\theta_0$ appropriately understood to be taken modulo $2\pi$. We illustrate this phase space in \cref{fig: phase space}, equivalent to the plot of \citet[Fig. 2b]{Zottl2013a}. In particular, it is helpful to emphasise the two distinct behavioural regimes on this timescale: (1) endless \emph{tumbling}, where the swimmer does not cross $z_0=0$ and $\theta_0$ varies monotonically, and (2) periodic \emph{swinging}, where the swimmer repeatedly crosses the midline of the flow and $\theta_0$ oscillates between two values, $\theta_0\in[\theta_{\text{min}},\theta_{\text{max}}]$, readily computed from \cref{eq: H_0(t)}. The separating trajectory passes through $(z_0,\theta_0)=(0,0)$ and corresponds to $H_0=g(0)$. Here, $H_0 > g(0)$ corresponds to tumbling, shaded grey in \cref{fig: phase space}, and $H_0 < g(0)$ corresponds to swinging. Of note, the period of these dynamics over the intermediate timescale, which we denote by $\tauperiod$, depends non-trivially on $H_0$.

We can identify these dynamics with those reported both numerically and experimentally by \citet{Omori2022}. In particular, as $H_0(t)$ approaches its minimum of $g(\pi)$, the trajectory in the phase space approaches a single point, $(z_0,\theta_0)=(0,\pi)$, corresponding to a swimmer that is directed upstream on the midline of the flow. This is precisely the so-called \emph{rheotactic} behaviour observed by \citet{Omori2022}, which they noted as the long-time behaviour of \cref{eq: original system} for particular definitions of $u(T)$ and $B(T)$, corresponding also to the experimentally determined behaviours of \textit{Chlamydomonas} in channel flow. This agreement suggests that the long-time dynamics of the full system 
may be captured by the evolution of the Hamiltonian-like quantity $H_0(t)$.

In order to examine this evolution, we consider the dynamics of the following Hamiltonian-like expression,
\begin{equation}\label{eq: H def}
    H(t) \coloneqq \frac{\gamma}{2\avg{u}}z^2 + g(\theta)\,.
\end{equation}
Taking the proper time derivative of \cref{eq: H def} and inserting \cref{eq: z-theta system} yields
\begin{equation}\label{eq: dHdt}
    \diff{H}{t} = \gamma\omega^{1/2} z \sin{\theta} \left[\frac{u(T)}{\avg{u}} - \frac{1 - B(T)\cos{2\theta}}{1 - \avg{B}\cos{2\theta}}\right]\,.
\end{equation}
Transforming the time derivative following \cref{eq: transform derivative}, we then insert our expansions of $z$ and $\theta$ into \cref{eq: dHdt}, noting that $H$ is equal to $H_0$ at leading order. Expanding $H\sim H_0 + \omega^{-1/2}H_1 + \omega^{-1}H_2+\cdots$, the balance at $\bigO{\omega}$ is simply $H_{0T}=0$, so we deduce that $H_0=H_0(\tau,t)$, as expected. At $\bigO{\omega^{1/2}}$, we have
\begin{equation}
    H_{1T} + H_{0\tau} = \gamma z_0\sin{\theta_0}\left[\frac{u(T)}{\avg{u}} - \frac{1 - B(T)\cos{2\theta_0}}{1 - \avg{B}\cos{2\theta_0}}\right]\,.
\end{equation}
Averaging over the fast timescale $T$, equivalent to applying the Fredholm solvability condition, we immediately see that the term in square brackets vanishes, so that $H_{0\tau} = 0$ and $H_0 = H_0(t)$ is also independent of $\tau$, as expected.

Finally, we consider the $\bigO{1}$ terms in \cref{eq: dHdt}, which may be stated as
\begin{equation}\label{eq: H: second order balance: z1 and theta1}
    H_{2T} + H_{1\tau} + H_{0t}  = h(T,\tau,t)\,,
\end{equation}
where
\begin{multline}\label{eq: h definition}
    h(T,\tau,t) \coloneqq \gamma (z_0\theta_1 + z_1\sin{\theta_0})\left[\frac{u(T)}{\avg{u}} - \frac{1 - B(T)\cos{2\theta_0}}{1 - \avg{B}\cos{2\theta_0}}\right]\\
     - \frac{2\gamma z_0\theta_1 \sin{\theta_0}\sin{2\theta_0}}{(1-\avg{B}\cos{2\theta_0})^2}\left[B(T) - \avg{B}\right]\,,
\end{multline}
and we note that the expressions in the square brackets each average to zero over a period in $T$. 
Inserting our expressions for $z_1$ and $\theta_1$ from \cref{eq: z1-theta1 solution}, we have
\begin{multline}\label{eq: long RHS of H PDE}
    h(T,\tau,t) = \gamma\left(\sin^2{\theta_0}\Iu(T) - \gamma z_0^2\cos{\theta_0}\cos{2\theta_0}\IB(T)\right)
    \left[\frac{u(T)}{\avg{u}} - \frac{1 - B(T)\cos{2\theta_0}}{1 - \avg{B}\cos{2\theta_0}}\right]\\
    + \frac{\gamma^2 z_0^2\sin{\theta_0}\sin{2\theta_0}\cos{2\theta_0}}{(1-\avg{B}\cos{2\theta_0})^2}\IB(T)(B(T) - \avg{B}) + [\cdot]\,,
\end{multline}
with $[\cdot]$ encompassing terms that have zero fast-timescale average, which here are those involving $\bar{z}_1$ and $\bar{\theta}_1$. It is also helpful to note that $2\IB(T)(B(T) - \avg{B})=(\IB^2)_T$, so that $\avg{\IB(B - \avg{B})}=0$ by the periodicity of $\IB$. Hence, the entire second line of \cref{eq: long RHS of H PDE} will vanish when averaged over a period in $T$. 
Averaging over $T$ and noting the further relations $\avg{\IB B}=\avg{\IB\avg{B}}$, $\avg{\Iu u}=\avg{\Iu\avg{u}}$, and
\begin{equation}
    (\Iu\IB)_T = \Iu(T)(B(T)-\avg{B}) + \IB(T)(u(T)-\avg{u})\,,
\end{equation}
the fast-timescale average of $h$ is simply
\begin{equation}\label{eq: avg of h}
    \avg{h} =  \gamma \cos{2\theta_0}\left(\frac{\gamma}{\avg{u}}z_0^2\cos{\theta_0} +\frac{\sin^2{\theta_0}}{1 - \avg{B}\cos{2\theta_0}}\right)
    \underbrace{\avg{\Iu(B - \avg{B})}}_{W}\,,
\end{equation}
where $W\coloneqq\avg{\Iu(B - \avg{B})}$ is constant. Thus, averaging \cref{eq: H: second order balance: z1 and theta1} over a period in $T$ and then over a period in $\tau$\footnote{Here, noting that $H_0$ is independent of $\tau$, we can naively average over a single oscillation in $\tau$, despite the $t$-dependence of $\tauperiod$, which would otherwise require the method of \citet{Kuzmak1959}.} yields the long-time evolution equation
\begin{equation}\label{eq: H0 evolution equation: generic}
    \diff{H_0}{t} = \gamma f(H_0) W\,,
\end{equation}
where
\begin{equation}\label{eq: f definition}
    f(H_0) \coloneqq \frac{1}{\tauperiod}\int_0^{\tauperiod} \cos{2\theta_0}\left(\frac{\gamma}{\avg{u}}z_0^2\cos{\theta_0} +\frac{\sin^2{\theta_0}}{1 - \avg{B}\cos{2\theta_0}}\right)\intd{\tau}\,,
\end{equation}
recalling that $\tauperiod$ denotes the period of the oscillatory dynamics in $\tau$ and $H_0$ is independent of $\tau$ and $T$. Notably, the integrand of \cref{eq: f definition} depends on the swimmer's speed and shape only via their fast-time averages $\avg{u}$ and $\avg{B}$. Therefore, all of the information encoding the dynamic variation of $u$ and $B$ about their mean is solely contained within the swimmer-dependent constant $W$. Of particular note, if either $u$ or $B$ is constant, then $W=0$ and, hence, $\mathrm{d}H_0/\mathrm{d}{t}=0$, so that there is no long-time drift of $H_0$. This analytically verifies the numerical observations of \citet{Omori2022}, who concluded that oscillations in both swimmer speed and shape were required to modify long-time behaviour.

\begin{figure}
    \centering
    \includegraphics[width=0.7\textwidth]{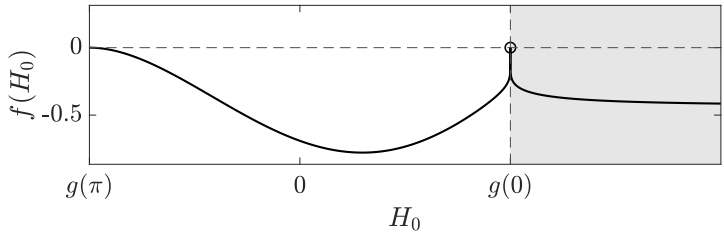}
    \caption{Exemplifying $f(H_0)$. We plot an example $f(H_0)$, as defined in \cref{eq: f definition} and computed numerically, for a range of $H_0$. The non-positivity of $f(H_0)$ is immediately evident, with $f\rightarrow0$ from below as $H_0\rightarrow g(\pi)$ or $H_0\rightarrow g(0)$. As noted in the main text, $f$ is undefined at $H_0=g(0)$, which we indicate with a hollow circle, but this point is readily seen to be half stable in the context of the dynamical system of \cref{eq: H0 evolution equation: generic}, so has negligible impact on the dynamics in practice. Here, we have fixed $\gamma=1$, $\avg{u}=1$, and $\avg{B}=0.5$. The shaded region corresponds to tumbling dynamics.}
    \label{fig: f}
\end{figure}

Having reduced the dynamics to the one-dimensional autonomous dynamical system of \cref{eq: H0 evolution equation: generic}, notably independent of $\omega$, it remains to understand $f(H_0)$, the average of a particular function of $z_0$ and $\theta_0$ over a period in $\tau$, which we illustrate in \cref{fig: f}. In \cref{app: deriving sign of f}, we analytically demonstrate that $f(H_0)\leq0$ for all $H_0$, in agreement with \cref{fig: f}. Hence, the sign of \cref{eq: H0 evolution equation: generic} is determined by $W$, which depends only on the dynamics of $u$ and $B$ over a single oscillation. Strictly, there is a higher-order problem to be solved close to $H_0=g(0)$, evidenced by the cusp-like profile in \cref{fig: f}, with $\tauperiod\rightarrow\infty$ and $f(H_0)\rightarrow 0$ as $H_0\rightarrow g(0)$. However, noting that $f(H_0)<0$ either side of $H_0=g(0)$, this point is half stable in the context of \cref{eq: H0 evolution equation: generic} \citep{Strogatz2018}, so that it is unstable in practice and does not materially impact on the evolving dynamics.

Thus, the fixed point at $H_0=g(\pi)$, corresponding to the rheotactic configuration $(z_0,\theta_0)=(0,\pi)$, is globally stable if $W>0$ and unstable if $W<0$. Hence, rheotaxis is the globally emergent behaviour at leading order if $W>0$, whilst endless tumbling arises if $W<0$.

\begin{figure}
    \centering
    \begin{overpic}[width=0.9\textwidth]{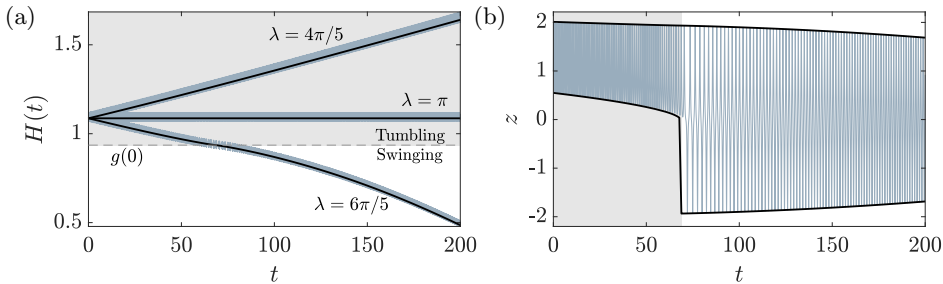}
    \put(-2,28){(a)}
    \put(49,28){(b)}
    \end{overpic}
    \caption{Numerical validation. (a) The value of $H$, as computed from the full numerical solution of \cref{eq: z-theta system} and the approximation of \cref{eq: H0 evolution equation: generic}, shown as blue and black curves, respectively, for three phase shifts $\lambda\in\{4\pi/5,\pi,6\pi/5\}$. (b) The asymptotically predicted bounds of $z$ oscillations for $\lambda=6\pi/5$ are shown as black curves, with the rapidly oscillating full solution shown in blue, highlighting excellent agreement even when the full solution transitions from tumbling dynamics towards rheotactic behaviour. Here, we have taken $(\alpha,\beta,\delta,\mu)=(1,0.5,0.32,0.3)$ and $\lambda\in\{4\pi/5,\pi,6\pi/5\}$ in the sinusoidal model of \citet{Omori2022}, fixing $\gamma=1$, $\omega=50$, and $(z,\theta)=(1,\pi/4)$ initially. The shaded regions correspond to tumbling dynamics.}
    \label{fig: examples}
\end{figure}

\section{Summary and conclusions}
Our analysis allows us to characterise the long-time behaviour of a swimmer in Poiseuille flow via the computation of a simple quantity, $W$, defined in \cref{eq: avg of h} and dependent only on the dynamics of the speed $u$ and the shape parameter $B$ of the swimmer over a single oscillation. In particular, we find that the long-time behaviours take one of three possible forms, given in terms of the leading-order Hamiltonian-like quantity $H_0$ of \cref{eq: H_0(t)}: (1) endless tumbling at increasing distance from the midline of the flow ($H_0\rightarrow\infty$); (2) preserved initial behaviour of the swimmer ($H_0 = H_0(0)$); (3) convergence to upstream rheotaxis, where the swimmer is situated at the midline of the flow ($H_0\rightarrow g(\pi)$). We find that the drift towards these long-time regimes is caused by the delicate higher-order interactions in the system. Specifically, the nonlinear interaction between the small  $\bigO{\omega^{-1/2}}$ variations from the leading-order system over the fast timescale $t = \bigO{\omega}$ gives rise to the significant $\bigO{1}$ effect over the slow timescale $t = \bigO{1}$.

In the context of \citeauthor{Omori2022}, where $u(T) = \alpha + \beta\sin{T}$ and $B(T)=\delta + \mu\sin{(T+\lambda)}$, we note that in-phase oscillations with $\lambda\in\{n\pi | n\in\mathbb{Z}\}$ immediately lead to $W=0$, corresponding to case (2) above. Any other values of $\lambda$ lead to $W\neq0$ (with maximal magnitude for $\lambda\in\{\pi/2 + n\pi | n\in\mathbb{Z}\}$) and long-term evolution of the swimmer behaviour. Notably, shifting $\lambda$ by $\pi$ results in a precise reversal of the sign of $W$ and a corresponding reversal of the sign of $\mathrm{d}H_0/\mathrm{d}t$. Hence, this shift in phase will precisely flip the fate of the swimmer, with rheotaxis being replaced by tumbling, or vice versa. Concretely, if $\beta\mu>0$, then $\lambda\in(0,\pi)$ results in tumbling, whilst $\lambda\in(\pi,2\pi)$ gives rise to rheotaxis. Further, our analysis predicts that swimmers having either $u$ or $B$ constant will not undergo a similar drift over $t=\bigO{1}$ at leading order. In particular, this highlights that rigid externally driven swimmers and Janus particles, associated with constant $B$, would differ fundamentally in their long-time behaviour from shape-changing swimmers with dynamically varying $B$.

In support of our asymptotic analysis, we present three numerical examples in \cref{fig: examples}, approximating $f(H_0)$ with quadrature using the easily obtained numerical solution of the simple ODE system of \cref{eq: first order solvability} in $\tau$ for fixed $H_0(t)$. In \cref{fig: examples}a, we compare the asymptotic and full numerical solutions through the evolution of $H$, adopting the sinusoidal model of \citet{Omori2022} and demonstrating excellent agreement between the solutions. This numerical validation spans the distinct dynamical regimes of tumbling and swinging, with different values of the phase shift $\lambda$ giving rise to distinct behaviours from otherwise identical initial conditions. \Cref{fig: examples}b captures a transition between behaviours, as observed by \citet{Omori2022} for $\lambda=6\pi/5$, where we have predicted the bounds of $z$ oscillations by combining the solution of \cref{eq: H0 evolution equation: generic} with \cref{eq: H_0(t)} evaluated at $\theta_0=\theta_{\text{min}}$ and $\theta_0=\pi$. We anticipate that similar simple calculations may be used to predict collisions between swimmers and channel boundaries in related experimental set-ups, though theoretical consideration of boundary effects in future work is warranted. More generally, studies of long-time swimmer dynamics in other scenarios, such as extensional flows, merit further investigation.

In summary, an asymptotic (three-timescale) multiple-scale analysis of the swimming model of \citet{Omori2022} has revealed a trichotomy of startlingly simple long-time behaviours, determined only by a single swimmer-specific constant of motion that may be readily computed a priori. This analysis confirms the numerical predictions of \citet{Omori2022}, in agreement with their experimental observations of \textit{Chlamydomonas}, and formally identifies the interacting oscillatory effects needed to elicit the eventual behaviours of endless tumbling and upstream rheotaxis in this model of swimming. 

B.J.W. is supported by the Royal Commission for the Exhibition of 1851. K.I. acknowledges JSPS-KAKENHI for Young Researchers (Grant No. 18K13456), JSPS-KAKENHI for Transformative Research Areas (Grant No. 21H05309) and JST, PRESTO, Japan (Grant No. JPMJPR1921). C.M. is supported by the Research Institute for Mathematical Sciences, an International Joint Usage/Research Center located at Kyoto University. M.P.D. is supported by the UK Engineering and Physical Sciences Research Council [Grant No. EP/W032317/1].

The computer code used in this study is available at \url{https://gitlab.com/bjwalker/emergent-rheotaxis-in-Poiseuille-flow}.

Declaration of Interests. The authors report no conflict of interest.

\appendix
\section{Deducing the sign of $f(H_0)$}\label{app: deriving sign of f}
Consider the integrand of \cref{eq: f definition}. Recalling the evolution equations of $z_0$ and $\theta_0$ on the intermediate timescale, given in \cref{eq: first order solvability}, this can be written as
\begin{equation*}
    \frac{\cos{2\theta_0}}{\avg{u}(1 - \avg{B}\cos{2\theta_0})}(z_0\sin{\theta_0})_{\tau}\,.
\end{equation*}
Integrating by parts then yields
\begin{equation}
    \tauperiod f(H_0) = \int_0^{\tauperiod}\frac{2z_0\sin{\theta_0}\sin{2\theta_0}}{\avg{u}(1-\avg{B}\cos{2\theta_0})^2} \theta_{0\tau}\intd{\tau} = \int\frac{2z_0\sin{\theta_0}\sin{2\theta_0}}{\avg{u}(1-\avg{B}\cos{2\theta_0})^2}\intd{\theta_0}\,,
\end{equation}
with boundary terms vanishing due to periodicity. With reference to the phase diagram of \cref{fig: phase space}, we note that we need only integrate in $\theta_0$ from its minimum attained value up to $\pi$, with both the integrand and the phase-space trajectory being symmetric about both $\theta_0=\pi$ and $z_0=0$. This corresponds to integrating over the branch of the trajectory in the upper-left quadrant of \cref{fig: phase space}, with the true value of $\tauperiod f(H_0)$ then being recovered by appropriate multiplication by two or four, depending on whether the trajectory is one of tumbling or swinging. Hence, we consider the integral only over the range $\theta_0\in[\theta_{\text{min}},\pi]$, where $\theta_{\text{min}}\in[0,\pi]$ is the minimum value attained by $\theta_0$ over an orbit, as specified solely by $H_0(t)$.

In the upper-left quadrant of the phase plane, $z_0$ is a non-negative increasing function of $\theta_0$, evident from \cref{eq: first order solvability: z0} and \cref{eq: H_0(t)}. In particular, $z_0(\theta_0 + \pi/2)\geq z_0(\pi/2)$ for $\theta_0\in[0,\pi/2]$. Further, the remaining combination of trigonometric terms in the integrand, denoted by $I(\theta_0)$ for brevity, satisfies $I(\theta_0)\geq0$ and $I(\theta_0 + \pi/2) = -I(\theta_0)$ for $\theta_0\in[0,\pi/2]$. Hence, the integral is trivially negative for $\theta_{\text{min}}\in[\pi/2,\pi]$, whilst for $\theta_{\text{min}}\in[0,\pi/2]$ we have
\begin{align}
    \int_{\theta_{\text{min}}}^{\pi} z_0(\theta_0)I(\theta_0)\intd{\theta_0} & = \int_{\theta_{\text{min}}}^{\pi/2} z_0(\theta_0)I(\theta_0)\intd{\theta_0} + \int_{\pi/2}^{\pi}z(\theta_0)I(\theta_0)\intd{\theta_0} \notag{}\\
    & \leq \int_{0}^{\pi/2} z_0(\pi/2)I(\theta_0)\intd{\theta_0} +  \int_0^{\pi/2}z_0(\theta_0 + \pi/2)I(\theta_0 + \pi/2)\intd{\theta_0} \notag{}\\
    & = \int_0^{\pi/2}[z_0(\pi/2) - z_0(\theta_0 + \pi/2)]I(\theta_0)\intd{\theta_0} \leq 0\,.
\end{align}
Hence, $\tauperiod f(H_0) \leq 0$, so that $f(H_0)\leq 0$ for all $H_0$. In particular, this equality is strict unless $H_0=g(\pi)$ or $H_0=g(0)$, which correspond to the degenerate rheotactic trajectory $(z_0,\theta_0)=(0,\pi)$ and the separating trajectory that lies between tumbling and swinging behaviours in \cref{fig: phase space}. As discussed in the main text, the case with $H_0=g(0)$ requires the consideration of a higher-order problem, though has negligible impact on the dynamics in practice.

\bibliography{references.bib}
\bibliographystyle{jfm}

\end{document}